\documentclass[twocolumn,letterpaper,aps,prl,superscriptaddress,showpacs,floatfix,preprintnumbers,longbibliography]{revtex4-1}

\usepackage{graphicx}   % Include figure filed
\usepackage{xspace}     % Include xspace
\usepackage{url}
\usepackage[mathlines]{lineno}% Enable numbering of text and display math
\usepackage{bm}
\usepackage{amsmath}
\usepackage{xcolor}
\usepackage[driverfallback=dvipdfm]{hyperref}
\hypersetup{
	colorlinks = true,
	citecolor= blue,
	linkbordercolor = cyan
}

\newcommand{\sqsn}{\mbox{$\sqrt{s_{_{NN}}}$}\xspace}

\newcommand{\lam}{\mbox{$\Lambda$}\xspace}
\newcommand{\alam}{\mbox{$\bar{\Lambda}$}\xspace}
\newcommand{\xin}{\mbox{$\Xi^{-}$}\xspace}
\newcommand{\xibar}{\mbox{$\bar{\Xi}^{+}$}\xspace}
\newcommand{\om}{\mbox{$\Omega^{-}$}\xspace}
\newcommand{\ombar}{\mbox{$\bar{\Omega}^{+}$}\xspace}

\newcommand{\mean}[1]{\langle #1 \rangle}

\def \be {\begin{equation}}
\def \ee {\end{equation}} 
\def \bea {\begin{eqnarray}}
\def \eea {\end{eqnarray}} 
\def \bP {{\bf P}}
\def \bs {{\bf s}}
\def \bv {{\bf v}}
\def \bomega {{\bm \omega}}

\def \curl {{\bm \nabla \times}}
\def \xim {\Xi^-}
\def \xip {\bar{\Xi}^+}
\def \snn {$\sqrt{s_{_{\rm NN}}}$} 
\usepackage{color}
\definecolor{orange}{cmyk}{0.,0.353,1.,0.}    % orange

\usepackage{lineno}
\newcommand*\patchAmsMathEnvironmentForLineno[1]{
  \expandafter\let\csname old#1\expandafter\endcsname\csname #1\endcsname
  \expandafter\let\csname oldend#1\expandafter\endcsname\csname end#1\endcsname
  \renewenvironment{#1}
     {\linenomath\csname old#1\endcsname}
     {\csname oldend#1\endcsname\endlinenomath}}
\newcommand*\patchBothAmsMathEnvironmentsForLineno[1]{
  \patchAmsMathEnvironmentForLineno{#1}
  \patchAmsMathEnvironmentForLineno{#1*}}
\AtBeginDocument{
\patchBothAmsMathEnvironmentsForLineno{equation}
\patchBothAmsMathEnvironmentsForLineno{align}
\patchBothAmsMathEnvironmentsForLineno{flalign}
\patchBothAmsMathEnvironmentsForLineno{alignat}
\patchBothAmsMathEnvironmentsForLineno{gather}
\patchBothAmsMathEnvironmentsForLineno{multline}
}

\begin{document}
%
% Should be commented out in the submission
%\preprint{\sl \new{version 5.1} \today. Text: \new{new}, \red{old}, \note{note}}

%Title of paper
\title{Global polarization of $\Xi$ and $\Omega$ hyperons 
in Au+Au collisions at \snn = 200 GeV}

%\author{T. Niida and S. A. Voloshin for the STAR Collaboration}
\affiliation{Abilene Christian University, Abilene, Texas   79699}
\affiliation{AGH University of Science and Technology, FPACS, Cracow 30-059, Poland}
\affiliation{Alikhanov Institute for Theoretical and Experimental Physics NRC "Kurchatov Institute", Moscow 117218, Russia}
\affiliation{Argonne National Laboratory, Argonne, Illinois 60439}
\affiliation{American University of Cairo, New Cairo 11835, New Cairo, Egypt}
\affiliation{Brookhaven National Laboratory, Upton, New York 11973}
\affiliation{University of California, Berkeley, California 94720}
\affiliation{University of California, Davis, California 95616}
\affiliation{University of California, Los Angeles, California 90095}
\affiliation{University of California, Riverside, California 92521}
\affiliation{Central China Normal University, Wuhan, Hubei 430079 }
\affiliation{University of Illinois at Chicago, Chicago, Illinois 60607}
\affiliation{Creighton University, Omaha, Nebraska 68178}
\affiliation{Czech Technical University in Prague, FNSPE, Prague 115 19, Czech Republic}
\affiliation{Technische Universit\"at Darmstadt, Darmstadt 64289, Germany}
\affiliation{ELTE E\"otv\"os Lor\'and University, Budapest, Hungary H-1117}
\affiliation{Frankfurt Institute for Advanced Studies FIAS, Frankfurt 60438, Germany}
\affiliation{Fudan University, Shanghai, 200433 }
\affiliation{University of Heidelberg, Heidelberg 69120, Germany }
\affiliation{University of Houston, Houston, Texas 77204}
\affiliation{Huzhou University, Huzhou, Zhejiang  313000}
\affiliation{Indian Institute of Science Education and Research (IISER), Berhampur 760010 , India}
\affiliation{Indian Institute of Science Education and Research (IISER) Tirupati, Tirupati 517507, India}
\affiliation{Indian Institute Technology, Patna, Bihar 801106, India}
\affiliation{Indiana University, Bloomington, Indiana 47408}
\affiliation{Institute of Modern Physics, Chinese Academy of Sciences, Lanzhou, Gansu 730000 }
\affiliation{University of Jammu, Jammu 180001, India}
\affiliation{Joint Institute for Nuclear Research, Dubna 141 980, Russia}
\affiliation{Kent State University, Kent, Ohio 44242}
\affiliation{University of Kentucky, Lexington, Kentucky 40506-0055}
\affiliation{Lawrence Berkeley National Laboratory, Berkeley, California 94720}
\affiliation{Lehigh University, Bethlehem, Pennsylvania 18015}
\affiliation{Max-Planck-Institut f\"ur Physik, Munich 80805, Germany}
\affiliation{Michigan State University, East Lansing, Michigan 48824}
\affiliation{National Research Nuclear University MEPhI, Moscow 115409, Russia}
\affiliation{National Institute of Science Education and Research, HBNI, Jatni 752050, India}
\affiliation{National Cheng Kung University, Tainan 70101 }
\affiliation{Nuclear Physics Institute of the CAS, Rez 250 68, Czech Republic}
\affiliation{Ohio State University, Columbus, Ohio 43210}
\affiliation{Institute of Nuclear Physics PAN, Cracow 31-342, Poland}
\affiliation{Panjab University, Chandigarh 160014, India}
\affiliation{Pennsylvania State University, University Park, Pennsylvania 16802}
\affiliation{NRC "Kurchatov Institute", Institute of High Energy Physics, Protvino 142281, Russia}
\affiliation{Purdue University, West Lafayette, Indiana 47907}
\affiliation{Rice University, Houston, Texas 77251}
\affiliation{Rutgers University, Piscataway, New Jersey 08854}
\affiliation{Universidade de S\~ao Paulo, S\~ao Paulo, Brazil 05314-970}
\affiliation{University of Science and Technology of China, Hefei, Anhui 230026}
\affiliation{Shandong University, Qingdao, Shandong 266237}
\affiliation{Shanghai Institute of Applied Physics, Chinese Academy of Sciences, Shanghai 201800}
\affiliation{Southern Connecticut State University, New Haven, Connecticut 06515}
\affiliation{State University of New York, Stony Brook, New York 11794}
\affiliation{Instituto de Alta Investigaci\'on, Universidad de Tarapac\'a, Arica 1000000, Chile}
\affiliation{Temple University, Philadelphia, Pennsylvania 19122}
\affiliation{Texas A\&M University, College Station, Texas 77843}
\affiliation{University of Texas, Austin, Texas 78712}
\affiliation{Tsinghua University, Beijing 100084}
\affiliation{University of Tsukuba, Tsukuba, Ibaraki 305-8571, Japan}
\affiliation{United States Naval Academy, Annapolis, Maryland 21402}
\affiliation{Valparaiso University, Valparaiso, Indiana 46383}
\affiliation{Variable Energy Cyclotron Centre, Kolkata 700064, India}
\affiliation{Warsaw University of Technology, Warsaw 00-661, Poland}
\affiliation{Wayne State University, Detroit, Michigan 48201}
\affiliation{Yale University, New Haven, Connecticut 06520}

\author{J.~Adam}\affiliation{Brookhaven National Laboratory, Upton, New York 11973}
\author{L.~Adamczyk}\affiliation{AGH University of Science and Technology, FPACS, Cracow 30-059, Poland}
\author{J.~R.~Adams}\affiliation{Ohio State University, Columbus, Ohio 43210}
\author{J.~K.~Adkins}\affiliation{University of Kentucky, Lexington, Kentucky 40506-0055}
\author{G.~Agakishiev}\affiliation{Joint Institute for Nuclear Research, Dubna 141 980, Russia}
\author{M.~M.~Aggarwal}\affiliation{Panjab University, Chandigarh 160014, India}
\author{Z.~Ahammed}\affiliation{Variable Energy Cyclotron Centre, Kolkata 700064, India}
\author{I.~Alekseev}\affiliation{Alikhanov Institute for Theoretical and Experimental Physics NRC "Kurchatov Institute", Moscow 117218, Russia}\affiliation{National Research Nuclear University MEPhI, Moscow 115409, Russia}
\author{D.~M.~Anderson}\affiliation{Texas A\&M University, College Station, Texas 77843}
\author{A.~Aparin}\affiliation{Joint Institute for Nuclear Research, Dubna 141 980, Russia}
% ...

\collaboration{STAR Collaboration}\noaffiliation

\date{\today}

\begin{abstract} %\linenumbers
Global polarization of $\Xi$ and $\Omega$ hyperons has been measured
for the first time in Au+Au collisions at $\sqrt{s_{_{NN}}}$ = 200
GeV. The measurements of the $\Xi^-$ and $\bar{\Xi}^+$ hyperon
polarization have been performed by two independent methods, via
analysis of the angular distribution of the daughter particles in the
parity violating weak decay $\Xi\rightarrow\Lambda+\pi$, as well as by
measuring the polarization of the daughter $\Lambda$-hyperon,
polarized via polarization transfer from its parent.  The
polarization, obtained by combining the results from the two methods
and averaged over $\Xi^-$ and $\bar{\Xi}^+$, is measured to be
$\langle P_\Xi \rangle = 0.47\pm0.10~({\rm stat.})\pm0.23~({\rm
  syst.})\,\%$ for the collision centrality 20\%-80\%.  The $\langle
P_\Xi \rangle$ is found to be slightly larger than the inclusive
$\Lambda$ polarization and in reasonable agreement with a multi-phase
transport model (AMPT).  The $\langle P_\Xi \rangle$ is found to
follow the centrality dependence of the vorticity predicted in the
model, increasing toward more peripheral collisions. The global
polarization of $\Omega$, $\langle P_\Omega \rangle =
1.11\pm0.87~({\rm stat.})\pm1.97~({\rm syst.})\,\%$ was obtained by
measuring the polarization of daughter $\Lambda$ in the decay $\Omega
\rightarrow \Lambda + K$, assuming the polarization transfer factor
$C_{\Omega\Lambda}=1$.
\end{abstract}

\maketitle

%\note{page break for word count} \clearpage

\setlength\linenumbersep{0.10cm}
%\linenumbers

%=======================================
% Introduction
%=======================================
%

The phenomenon of global polarization in heavy-ion collisions arises
from the partial conversion of the orbital angular momentum of
colliding nuclei into the spin angular momentum of the particles
produced in the
collision~\cite{Liang:2004ph,Voloshin:2004ha,Becattini_2008}.  As a
result, these particles become globally polarized along the direction
of the initial orbital momentum of the nuclei.  Global polarization
was first observed by the STAR Collaboration in the beam energy scan
Au+Au collisions~\cite{polBES} and was later confirmed, to better
precision, in the analysis of the 200~GeV data with high
statistics~\cite{Adam:2018ivw}. Assuming local thermal equilibrium,
the polarization of the produced particles is determined by the local
thermal vorticity of the fluid~\cite{Becattini_2008}. In the
nonrelativistic limit (for hyperons $m_H \gg T$, where $T$ is the
temperature), the polarization of the particles is given
by~\cite{Becattini:2016gvu}:
\be
\bP=\frac{\mean{\bs}}{s} \approx \frac{(s+1)}{3}\frac{\bomega}{T},
\label{eq:polar}
\ee
where $s$ is the spin of the particle, $\mean{\bs}$ is the mean spin
vector, and $\bomega=\frac{1}{2}\curl \bv$ is the local vorticity of
the fluid velocity field.  Averaged over the entire system volume, the
vorticity direction should coincide with the direction of the system
orbital momentum.

Following from Eq.~\ref{eq:polar}, all particles, as well as
antiparticles of the same spin should have the same polarization.  A
difference could arise from effects of the initial magnetic
field~\cite{Becattini:2016gvu}, from the fact that different particles
are produced at different times or regions as the system freezes
out~\cite{Vitiuk:2019rfv}, or through meson-baryon
interactions~\cite{Csernai:2018yok}.  Thus far, only $\lam$ and
$\alam$ polarizations have been
measured~\cite{polBES,Adam:2018ivw,Acharya:2019ryw}.  Therefore, to
establish the global nature of the polarization, it is very important
to measure the polarization of different particles, and if possible,
particles of different spins.  In the global polarization picture based
on vorticity one expects different particles to be polarized in the
same direction and that the polarization magnitudes for different
particles depend only on their spin in accordance to
Eq.~\ref{eq:polar}.

In order to study the possible contribution from the initial magnetic
field, the polarization measurement with particles of different
magnetic moment would provide additional information.  The difference
in the polarization measured so far between $\lam$ and $\alam$ is not
significant and is at the level of a couple standard deviations at
most.

Although the energy dependence of the average \lam polarization can be
explained well by theoretical
models~\cite{polHydro,polAMPT,Sun:2017xhx,Xie:2017upb,Ivanov:2019ern,Vitiuk:2019rfv},
many questions remain open, and the detail modeling of the global
polarization and dynamical treatment of spin are under development.
In fact, there exist sign problems in differential measurements of the
global and local polarizations, not only between the experimental data
and models but also among different
models~\cite{Adam:2019srw,Niida:2018hfw,Becattini:2020ngo}.  
For example, \lam (\alam) polarization along the beam
  direction measured experimentally~\cite{Adam:2019srw} 
  differ in the sign and magnitude of the effect from many theoretical
  calculations.
Therefore, further experimental inputs are crucial for understanding
the vorticity and polarization phenomena in heavy-ion collisions.  In
this paper we present the first measurements of the global
polarization of spin $s=1/2$~~ $\xim$ and $\xip$ hyperons, as well as
spin $s=3/2$~ $\Omega$ hyperons in Au+Au collisions at \snn = 200 GeV.

Hyperon weak decays present the most straightforward possibility for
measuring the polarization of the produced particles~\cite{Bunce:1976yb}. 
In parity-violating weak decays the daughter particle distribution 
in the rest frame of the hyperon directly depends on the hyperon polarization:
\be
\frac{dN}{d\Omega^{\ast}} =\frac{1}{4\pi} \left(1 
+ \alpha_{H} \bP_{H}^{\ast}\cdot \hat{\bm p}_B^* \right), 
\label{eq:hyperon_decay}
\ee
where $\alpha_H$ is the hyperon decay parameter, $\bP_H^*$ is the
hyperon polarization, and $\hat{\bm p}_B^*$ is the unit vector in the
direction of the daughter baryon momentum, both in the parent rest
frame denoted by an asterisk.

$\xim$ ($\xip$) hyperon decay happens in two steps: $\xim \rightarrow
\Lambda +\pi^-$ with subsequent decay $\Lambda \rightarrow p+
\pi^-$. If $\xim$ is polarized, its polarization is partially
transferred to the daughter $\Lambda$.  Both steps in such a cascade decay
are parity violating and thus can be used for an independent
measurement of the polarization of $\xim$ ($\xip$).

%-----------------------------------------------------------
The polarization of the daughter baryon in a weak decay of a spin 1/2
hyperon is described by the Lee-Yang
formula~\cite{Lee:1957qs,Luk:2000zw,Huang:2004jp} in terms of the three
parameters $\alpha$ (parity violating part), $\beta$ (violation of the
time reversal symmetry), and $\gamma$ (satisfying $\alpha^2+\beta^2+\gamma^2=1$). 
For a particular case of
$\Xi\rightarrow \Lambda + \pi$ decay it reads:
%
%\begin{widetext}
\begin{equation}\label{eq:poldecay}
{\bf P}^*_{\Lambda}=
\frac{(\alpha_{\Xi}+{\bf P}^*_{\Xi}\cdot\hat{\bm p}_\Lambda^*)\hat{\bm p}_\Lambda^* 
  + \beta_{\Xi}{\bf P}^*_{\Xi}\times\hat{\bm p}_\Lambda^* 
    + \gamma_{\Xi} \hat{\bm p}_\Lambda^*\times({\bf P}^*_{\Xi}\times\hat{\bm
    p}_\Lambda^*)}{1+\alpha_{\Xi}{\bf P}^*_{\Xi} \cdot\hat{\bm p}_\Lambda^*} ,
\end{equation}
%\end{widetext}
%
where $\hat{\bm p}_\Lambda^*$ is the unit vector of the $\Lambda$ momentum
in the $\Xi$ rest frame. Averaging over the angular distribution of
the $\Lambda$ in the rest frame of the $\Xi$ given by
Eq.~\ref{eq:hyperon_decay} yields
\be
{\bf  P}^*_{\Lambda} = C_{\Xi^- \Lambda} {\bf P}^*_{\Xi} =
\tfrac{1}{3} \left( 1+2\gamma_{\Xi} \right) {\bf P}^*_{\Xi}.  \label{eq:PlamPxi}
\ee
Using the measured value for the $\gamma_{\Xi}$ 
parameter~\cite{Zyla:2020zbs,Huang:2004jp}, 
the polarization transfer coefficient for $\Xi^{-}$ to $\Lambda$ decay is:
\begin{align}
C_{\Xi^- \Lambda}=&\tfrac{1}{3}\left(1+2\times0.916\right)=+0.944. 
%\nonumber \\ 
%&\tfrac{1}{3}\left(2\times0.85+1\right)=+0.900
\end{align}

The polarization of the daughter baryon in a two particle decay of
spin $3/2$ hyperon, $\Omega\rightarrow \Lambda +K$, is also described
by three parameters $\alpha_\Omega$, $\beta_\Omega$, and
$\gamma_\Omega$~\cite{Luk:1988as}.  The decay parameter
$\alpha_\Omega$, determines the angular distribution of \lam in the
$\Omega$ rest frame and is measured to be small~\cite{Zyla:2020zbs}:
$\alpha_\Omega=0.0157\pm0.0021$; this makes it practically impossible
to measure the $\Omega$ polarization via analysis of the daughter
$\Lambda$ angular distribution. The polarization transfer in this case
is determined by the $\gamma_\Omega$ parameter
via~\cite{Luk:1983pe,Luk:1988as,Kim:1992az}:
\be 
{\bf  P}^*_{\Lambda} = C_{\Omega^- \Lambda} {\bf P}^*_{\Omega} =
\tfrac{1}{5} \left( 1+4\gamma_{\Omega} \right) {\bf P}^*_{\Omega}.  
\label{eq:PlamPom}
\ee
The time-reversal violation parameter $\beta_\Omega $ is expected to
be small. This combined with the constraint that
$\alpha^2+\beta^2+\gamma^2=1$ limits the unmeasured parameter to
$\gamma_\Omega \approx \pm 1$, resulting in a polarization transfer
$C_{\Omega^- \Lambda} \approx 1$ or $C_{\Omega^- \Lambda}\approx
-0.6$.

%=======================================
% STAR detector and Dataset
%=======================================
Our analysis is based on the data of Au+Au collisions at \sqsn =
200~GeV collected in 2010, 2011, 2014, and 2016 by the STAR
detector. Charged-particle tracks were measured in the time projection
chamber (TPC)~\cite{tpc}, which covers the full azimuth and a
pseudorapidity range of $|\eta|<1$.  The collision vertices were
reconstructed using the measured charged-particle tracks and were
required to be within 30~cm relative to the TPC center in the beam
direction for the 2010 and 2011 datasets to ensure a good acceptance
of reconstructed tracks.  The narrower vertex selection to be within
6~cm was applied in the 2014 and 2016 data due to online trigger
requirement for the Heavy Flavor Tracker (HFT) installed prior to 2014 data
taking.  The vertex in the radial direction relative to the beam
center was also required to be within 2~cm to reject background from
collisions with beam pipe.  Additionally, the difference in the vertex
positions along the beam direction from the vertex position detectors
(VPD)~\cite{vpd} located at forward and backward pseudorapidities
($4.24<|\eta|<5.1$) was required to be less than 3~cm to suppress
pileup events in which more than one heavy-ion collision occurred.
These selection criteria yielded about 180 (350) million minimum bias
(MB) events for the 2010 (2011) dataset, 1 billion MB events for the
2014 dataset, and 1.5 billion MB events for the 2016 dataset.  The MB
trigger requires hits of both VPDs and the zero-degree calorimeters
(ZDCs)~\cite{zdc}, which detect spectator neutrons in $|\eta|>6.3$.
The collision centrality was determined from the measured multiplicity
of charged particles within $|\eta|<0.5$ and a Monte Carlo Glauber
simulation~\cite{glauber,BESv2}.

The first-harmonic event plane angle $\Psi_1$ as an experimental
estimate of the impact parameter direction was determined by measuring
the neutron spectator deflection~\cite{Voloshin:2016ppr} in the ZDCs
equipped with Shower Maximum Detectors
(SMD)~\cite{SN0448,v1smd,Adamczyk:2017ird}.  The event plane
resolution~\cite{TwoSub} is largest ($\sim$41\%, the resolution is
better if it is closer to 100\%) at 30\%-40\% collision centrality for
the 2014 and 2016 datasets, and is decreased by 4\% for the 2010 and
2011 datasets~\cite{Adam:2018ivw}.

%=======================================
% Fig.1  
%=======================================
\begin{figure}[hbt]
\begin{center}
\includegraphics[width=\linewidth, bb=0 0 567 291]{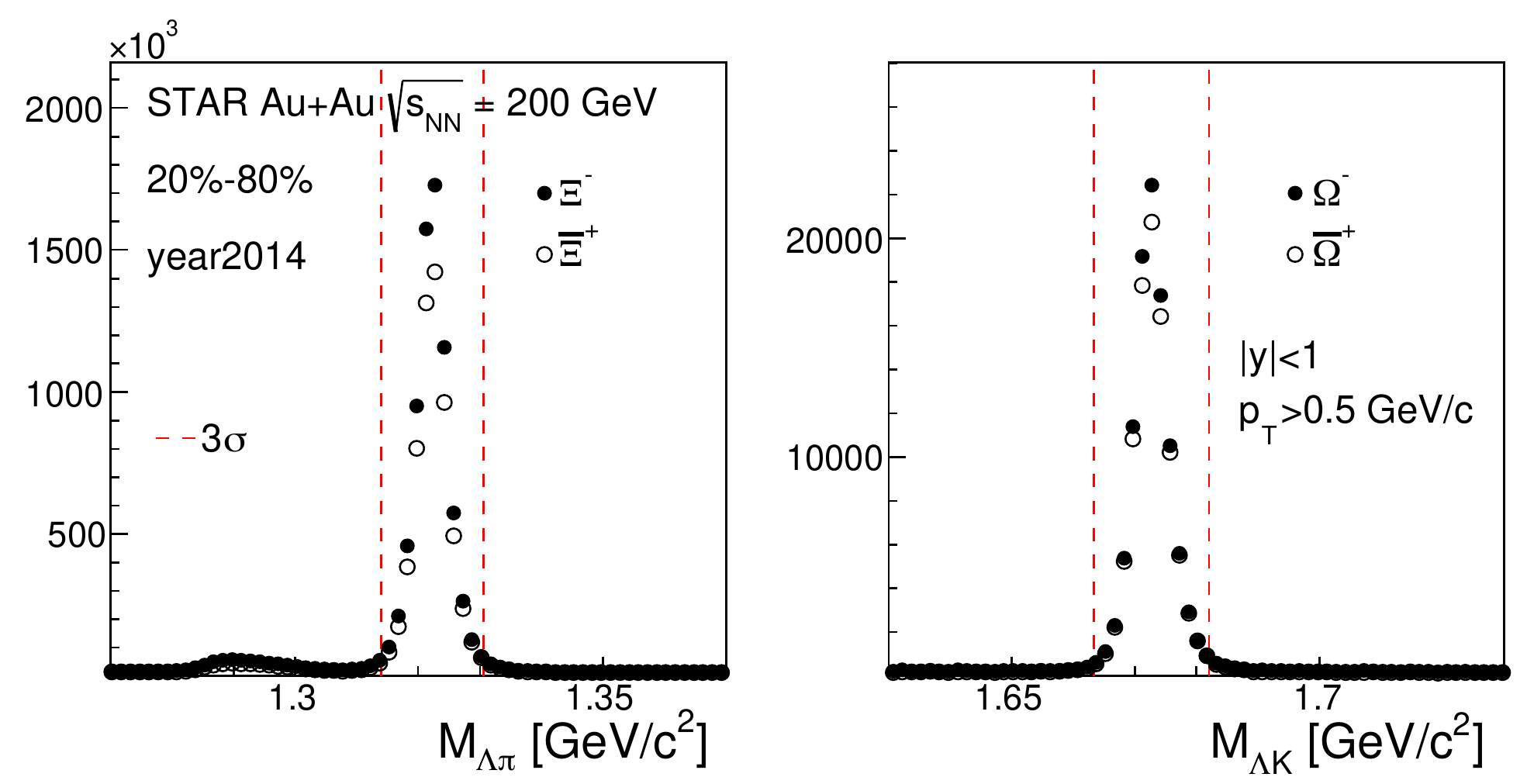}
\caption{\label{fig:mass}(Color online) Invariant mass distributions
  of \xin (\xibar) and \om (\ombar) for 20\%-80\% centrality in Au+Au
  collisions at \sqsn = 200~GeV taken in 2014. Vertical dashed lines
  indicate three standard deviations (3$\sigma$) from the peak
  positions, assuming a normal distribution.}
\end{center}
\end{figure}
%=======================================

The parent \xin (\xibar), \om (\ombar), and their daughter \lam
(\alam) were reconstructed utilizing the decay channels of
$\Xi^{-}\rightarrow\Lambda\pi^{-}$ (99.887\%),
$\Omega^{-}\rightarrow\Lambda K^{-}$ (67.8\%), and $\Lambda\rightarrow
p\pi^{-}$ (63.9\%), where the numbers in parenthesis indicate the
corresponding branching ratio of the decays~\cite{Zyla:2020zbs}.
Charged pions (kaons) and protons of the daughter particles were
identified based on the ionization energy loss in the TPC gas, and the
timing information measured by the Time-Of-Flight
detector~\cite{tof}. Reconstruction of \xin (\xibar), \om (\ombar), and
\lam (\alam) was performed using the KF Particle Finder package based
on the Kalman Filter (KF) method initially developed for the CBM and ALICE
experiments~\cite{Gorbunov,Zyzak,Kisel:2018nvd}, which utilizes the
quality of the track fit as well as the decay topology.
Figure~\ref{fig:mass} shows the invariant mass distributions for
reconstructed \xin (\xibar) and \om (\ombar) for 20\%-80\%
centrality. The purities for this centrality bin are higher than 90\%
for both species. The significance with the Kalman Filter method is
found to be increased by $\sim30\%$ for $\Xi$ compared to the
traditional method for reconstruction of short-lived particles
(e.g. see Refs.~\cite{Adam:2019koz,Adam:2018ivw}).  The hyperon
candidates were also ensured not to share their decay products with
other particles of interest.

%The polarization projected along the initial angular momentum direction $\hat{{\it J}}$ can be
The polarization along the initial angular momentum direction can be
defined as~\cite{pol2007}:
\begin{equation}
%P_H = \langle {\bf P}^*_H \cdot \hat{{\it J}} \rangle = \frac{8}{\pi\alpha_H}\frac{\langle\sin(\Psi_1^{\rm obs}-\phi_B^\ast)\rangle}{\rm Res(\Psi_1)},
P_H = \frac{8}{\pi\alpha_H}\frac{\langle\sin(\Psi_1^{\rm obs}-\phi_B^\ast)\rangle}{\rm Res(\Psi_1)},
\end{equation}
where $\alpha_H$ is the hyperon decay parameter and $\phi_B^\ast$ is the
azimuthal angle of the daughter baryon in the parent hyperon rest frame. The
azimuthal angle of the first-order event plane is $\Psi_1^{\rm obs}$, and
Res($\Psi_1$) is the resolution~\cite{TwoSub} with which it estimates the reaction plane.

The extraction of $\langle\sin(\Psi_1^{\rm obs}-\phi^\ast)\rangle$ was
performed in the same way as in our previous
studies~\cite{polBES,Adam:2018ivw}. The decay parameters of \lam,
\xin, and \om have been recently updated by the Particle Data
Group~\cite{Zyla:2020zbs} and the latest values are used in this
analysis; $\alpha_\Lambda=0.732\pm0.014$, $\alpha_\Xi=-0.401\pm0.010$,
and $\alpha_\Omega=0.0157\pm0.0021$.  When comparing to earlier
measurements, the previous results are rescaled by using the new
values, i.e. $\alpha_{\rm old}/\alpha_{\rm new}$.  In case of the
$\Xi$ and $\Omega$ hyperon polarization measurements via measurements
of the daughter \lam polarization, the polarization transfer factors
$C_{\Xi\Lambda (\Omega\Lambda)}$ from Eqs.~\ref{eq:PlamPxi} and
\ref{eq:PlamPom} are used to obtain the parent polarization.

The largest systematic uncertainty (37\%) was attributed to the
variation of the results obtained with datasets taken in different
years. The difference could be partly due to the change in the
detector configuration (inclusion of the HFT in the 2014 and 2016 data
taking) and increased luminosity in recent years, both of which lead
to the reduction of detecting efficiency. After careful checks of the
detector performance and detailed quality assurance of the data,
weighted average over different datasets was used as the final
result. All other systematic uncertainties were assessed based on the
weighted average: by comparing different polarization signal
extractions~\cite{Adam:2018ivw} (11\%), by varying the mass window for
particles of interest from 3$\sigma$ to 2$\sigma$ (15\%), by varying
the decay lengths of both parent and daughter hyperons (4\%), and by
considering uncertainties on the decay parameter $\alpha_H$ (2\%),
where the numbers in parentheses represent the uncertainty for the
$\Xi$ polarization via the daughter \lam polarization measurement.  A
correction for non-uniform acceptance effects~\cite{pol2007} was
applied for the appropriate detector configuration for the given
dataset.  This correction, depending on particle species, was less
than 2\%.  Due to a weak $p_T$ dependence on the global
polarization~\cite{Adam:2018ivw}, effects from the $p_T$ dependent
efficiency of the hyperon reconstruction were found to be negligible.

%=======================================
% Fig.2  
%=======================================
\begin{figure}[hbt]
\begin{center}
\includegraphics[width=\linewidth,bb=0 0 567 431]{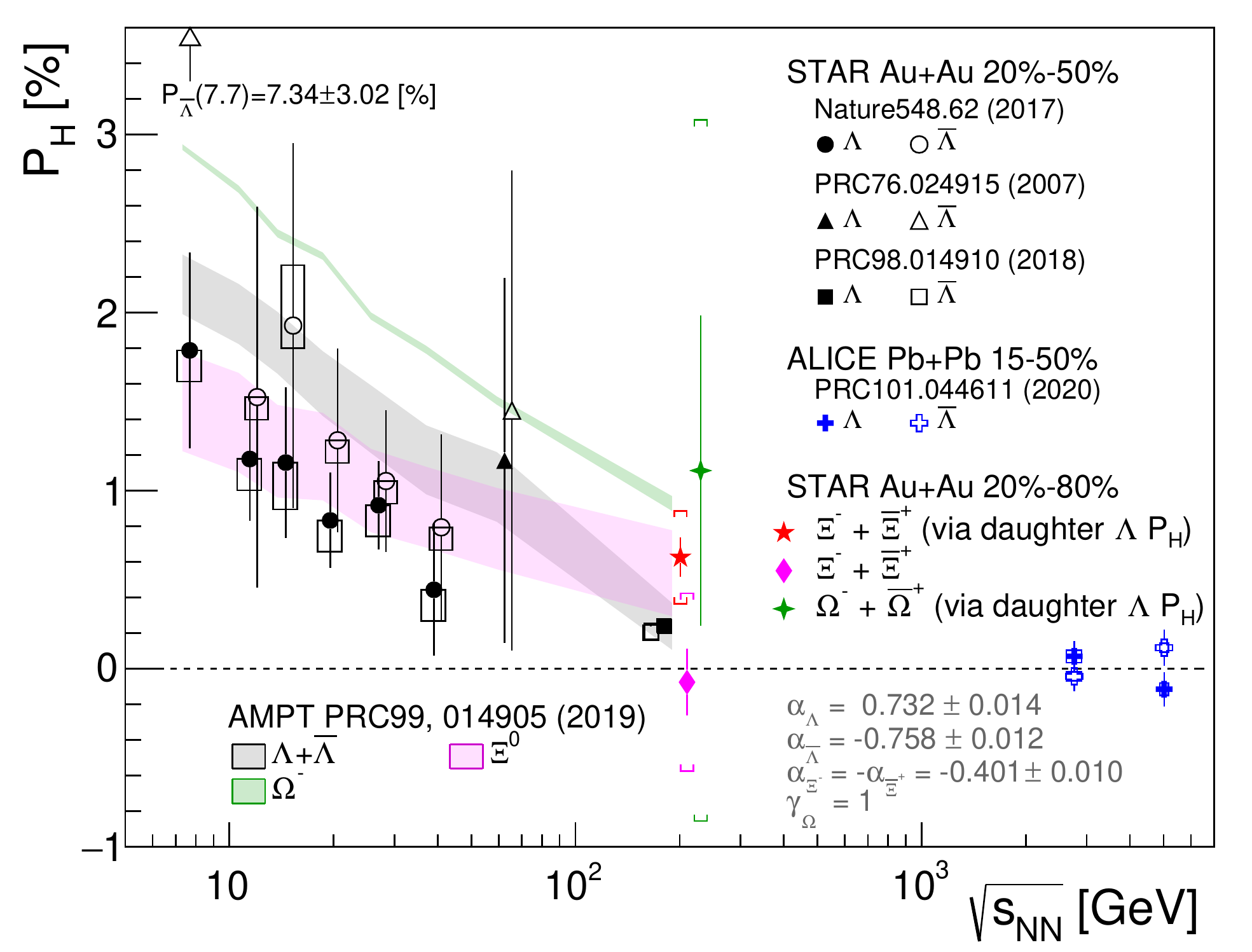} % for PRL
\caption{ (Color online) The energy dependence of the hyperon global
  polarization measurements.  The points corresponding to $\lam$ and
  $\alam$ polarizations, as well as $\Xi$ and $\Omega$ points in Au+Au
  collisions at \snn~=~200~GeV are slightly shifted for clarity. 
  Previous results from the STAR~\cite{polBES,pol2007,Adam:2018ivw} and 
  ALICE~\cite{Acharya:2019ryw} experiments compared here are rescaled 
  by new decay parameter indicated inside the figure.
  The data point for $\alam$ at 7.7 GeV is out of the axis range and indicated 
  by an arrow with the value. The results of the AMPT model 
  calculations~\cite{Wei:2018zfb} for 20-50\% centrality are shown by shaded bands
  where the band width corresponds to the uncertainty of the calculations. 
  %The model calculation uncertainties at \snn~=~200~GeV
  %are similar to the ones at lower energy and are not shown for clarity.  
}
\label{fig:gpolsnn}
\end{center}
\end{figure}
%=======================================

Figure~\ref{fig:gpolsnn} shows the collision energy dependence of the
\lam hyperon global polarization measured
earlier~\cite{polBES,pol2007,Adam:2018ivw,Acharya:2019ryw} together
with the new results on $\Xi$ and $\Omega$ global polarizations at
\sqsn~=~200~GeV. (Note that the statistical and systematic
uncertainties for the \lam~ are smaller than the symbol size.)  For
both $\Xi$ and $\Omega$ polarizations, the particle and antiparticle
results are averaged to reduce the statistical uncertainty. Also to
maximize the significance of the polarization signal, the results were
integrated over the centrality range 20\%-80\%, transverse momentum
$p_T>0.5$~GeV/$c$, and rapidity $|y|<1$.  Global polarization of
$\xin$ and $ \xibar$ measurements via daughter \lam polarization show
positive values, with no significant difference between \xin and
\xibar ($P_\Xi~(\%)=0.77\pm0.16~({\rm stat.})\pm0.49~({\rm syst.})$
and $P_{\bar{\Xi}}~(\%)=0.49\pm0.16~({\rm stat.})\pm0.20~({\rm
  syst.})$).  The average polarization value obtained by this method
is $\langle P_\Xi\rangle~(\%)=0.63\pm0.11~({\rm stat.})\pm0.26~({\rm
  syst.})$.  The $\Xi+\bar{\Xi}$ polarization was also measured via
analysis of the angular distribution of daughter \lam in $\Xi$ rest
frame. This result, $\langle P_\Xi\rangle~(\%)=-0.07\pm0.19~({\rm
  stat.})\pm0.50~({\rm syst.})$, has larger uncertainty in part due to
a smaller value of $\alpha_\Xi$ compared to $\alpha_\Lambda$, which
leads to smaller sensitivity of the measurement.  
Note that with given uncertainties the difference between 
the two methods is within 1$\sigma$.
The weighted average of the two measurements is $\langle
P_\Xi\rangle~(\%)=0.47\pm0.10~({\rm stat.})\pm0.23~({\rm syst.})$,
which is larger than the polarization of inclusive \lam+\alam measured
at the same energy for 20\%-80\% centrality, $\langle
P_\Lambda\rangle~(\%)=0.24\pm0.03\pm0.03$~\cite{Adam:2018ivw},
although the difference is still not significant considering the
statistical and systematic uncertainties of both measurements.  Note
that the above quoted values for the inclusive \lam have been rescaled
by the new decay parameter as mentioned earlier and 
``inclusive" means \lam coming from primary vertex as well as 
those decaying from higher states.

Calculations~\cite{Wei:2018zfb} carried out with a multi-phase
transport model (AMPT) can describe the particle species dependence in
data at 200 GeV as well as the energy dependence for $\Lambda$. These
calculations indicate that the lighter particles with higher spin
could be more polarized by the vorticity~\cite{Wei:2018zfb}. 
The multi-strange particles might freeze out at
earlier times, which may lead to larger polarization for $\Xi$ and
$\Omega$ compared to $\Lambda$~\cite{Vitiuk:2019rfv}.  
The feed-down effect can also lead to
a $15\sim20\%$ reduction of the primary \lam
polarization~\cite{polHydro,Becattini:2016gvu,polAMPT,Xia:2019fjf},
while the $\Xi$ has less contribution from the feed-down. All these
effects can contribute to small differences in the measured
polarizations between inclusive \lam and $\Xi$ hyperons.

Global polarization of \om was also measured and is presented in
Fig.~\ref{fig:gpolsnn} under the assumption of $\gamma_\Omega=+1$ and
therefore $C_{\Omega\Lambda}=1$, as discussed with respect to
Eq.~\ref{eq:PlamPom}.  The result has large uncertainty, $\langle
P_\Omega\rangle~(\%)=1.11\pm0.87~({\rm stat.})\pm1.97~({\rm syst.})$
for 20\%-80\% centrality.
%but the mean value seems to follow the particle species dependence predicted by the AMPT model.  
Assumption of $\gamma_\Omega=-1$ (therefore $C_{\Omega\Lambda}=-0.6$)
results in $\langle P_\Omega\rangle~(\%)=-0.67\pm0.52~({\rm
  stat.})\pm1.18~({\rm syst.})$.  Assuming the validity of the global
polarization picture, $\langle P_{\Omega}\rangle$ should be positive, and therefore 
the result favors $\gamma_\Omega\approx +1$
instead of $\gamma_\Omega\approx-1$, but the uncertainties are large
and more precise measurements are needed to make a definitive
statement.

% Fig.3 
%=======================================
\begin{figure}[t]
\begin{center}
\includegraphics[width=\linewidth,bb=0 0 567 534]{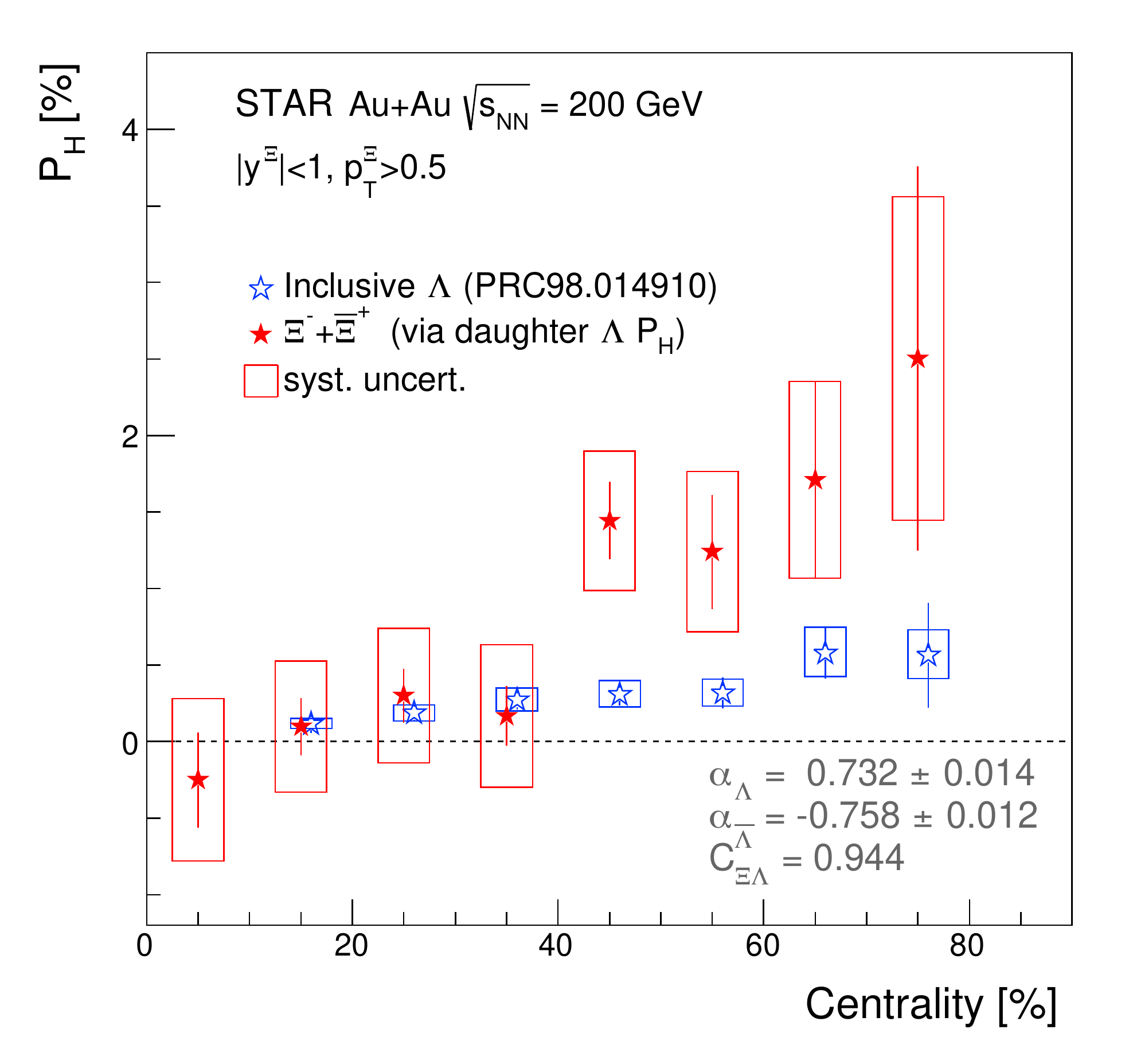}
\caption{ (Color online) The global polarization of $\Xi$ hyperons
  obtained via measurements of the polarization of daughter $\Lambda$
  hyperons as a function of the collision centrality in Au+Au
  collisions at \sqsn = 200 GeV.  Open boxes show the systematic
  uncertainties. Results for the inclusive $\Lambda$
  measurements~\cite{Adam:2018ivw} are shown for comparison.  }
\label{fig:PHcent}
\end{center}
\end{figure}
%=======================================

The centrality dependence of $\Xi+\bar{\Xi}$ polarization via the
measurement of daughter \lam polarization is shown in
Fig.~\ref{fig:PHcent}, where the inclusive \lam
polarization~\cite{Adam:2018ivw} is plotted for comparison.  The
hyperon polarization increases in more peripheral collisions as
expected from the centrality dependence of the fluid
vorticity~\cite{Jiang:2016woz,Xie:2017upb}.  The $\Xi$ polarization
looks larger than that of the inclusive \lam in peripheral collisions
as already discussed in relation to Fig.~\ref{fig:gpolsnn}, although
the uncertainties preclude a more definite conclusion.

In summary, we have presented the first measurements of the global
polarization for \xin (\xibar) hyperons in Au+Au collisions at \sqsn =
200 GeV. Our results of $\Xi$ hyperon polarization, along with the
previous measurements of \lam polarization, confirm the global
polarization picture based on the system fluid vorticity. The average
polarization of $\Xi+\bar{\Xi}$ seems to be larger than that of the
inclusive \lam, which is qualitatively captured by the AMPT model.
The measured polarization seems to exhibit a centrality dependence as
expected from the impact parameter dependence of the vorticity. Global
polarization of \om hyperons was, also for the first time, extracted
via measurements of the polarization of the daughter \lam and
presented with the assumption that $\gamma_\Omega=+1$. Future
measurements with higher precision will shed light on the uncertainty
of the decay parameter $\gamma_\Omega$, as well as experimental
results on the global polarization of spin-3/2 particles, providing
critical information about spin dynamics in heavy-ion collisions.

%
%\note{***
%  page break for PRL word count $<$3.5 pages}
%\clearpage
%
%
%
%
%%%%%%%%%%%%%%%%%%%%%%%%%  Acknowledgements 
\begin{acknowledgments}
We thank the RHIC Operations Group and RCF at BNL, the NERSC Center at LBNL, and the Open Science Grid consortium for providing resources and support.  This work was supported in part by the Office of Nuclear Physics within the U.S. DOE Office of Science, the U.S. National Science Foundation, the Ministry of Education and Science of the Russian Federation, National Natural Science Foundation of China, Chinese Academy of Science, the Ministry of Science and Technology of China and the Chinese Ministry of Education, the Higher Education Sprout Project by Ministry of Education at NCKU, the National Research Foundation of Korea, Czech Science Foundation and Ministry of Education, Youth and Sports of the Czech Republic, Hungarian National Research, Development and Innovation Office, New National Excellency Programme of the Hungarian Ministry of Human Capacities, Department of Atomic Energy and Department of Science and Technology of the Government of India, the National Science Centre of Poland, the Ministry  of Science, Education and Sports of the Republic of Croatia, RosAtom of Russia and German Bundesministerium fur Bildung, Wissenschaft, Forschung and Technologie (BMBF), Helmholtz Association, Ministry of Education, Culture, Sports, Science, and Technology (MEXT) and Japan Society for the Promotion of Science (JSPS).

%We thank the RHIC Operations Group and RCF at BNL, the NERSC Center at
%LBNL, and the Open Science Grid consortium for providing resources and
%support.  This work was supported in part by the Office of Nuclear
%Physics within the U.S. DOE Office of Science, the U.S. National
%Science Foundation, the Ministry of Education and Science of the
%Russian Federation, National Natural Science Foundation of China,
%Chinese Academy of Science, the Ministry of Science and Technology of
%China and the Chinese Ministry of Education, the Higher Education
%Sprout Project by Ministry of Education at NCKU, the National Research
%Foundation of Korea, Czech Science Foundation and Ministry of
%Education, Youth and Sports of the Czech Republic, Hungarian National
%Research, Development and Innovation Office, New National Excellency
%Programme of the Hungarian Ministry of Human Capacities, Department of
%Atomic Energy and Department of Science and Technology of the
%Government of India, the National Science Centre of Poland, the
%Ministry of Science, Education and Sports of the Republic of Croatia,
%RosAtom of Russia and German Bundesministerium fur Bildung,
%Wissenschaft, Forschung and Technologie (BMBF), Helmholtz Association,
%Ministry of Education, Culture, Sports, Science, and Technology (MEXT)
%and Japan Society for the Promotion of Science (JSPS).
\end{acknowledgments}
%
%%%%%%%%%%%%%%%%%%%%%%%%%%%  References 
% BibTeX:
\bibliography{ref_xipol}   
\end{document}